# Relativistic study of the energy-dependent Coulomb potential including Coulomb-like tensor interaction

M. Hamzavi[1*], S. M. Ikhdair[2**]

[1]*Department of Basic Sciences, Shahrood Branch, Islamic Azad University, Shahrood, Iran*
[2]*Physics Department, Near East University, 922022 Nicosia, North Cyprus, Mersin 10, Turkey*

[*] Corresponding author: *majid.hamzavi@gmail.com*
*Tel.:+98 273 3395270, fax: +98 273 3395270*

[**] *sikhdair@neu.edu.tr*

**Abstract**

The exact Dirac equation for the energy-dependent Coulomb (EDC) potential including a Coulomb-like tensor (CLT) potential has been studied in the presence of spin and pseudospin (p-spin) symmetries with arbitrary spin-orbit quantum number $\kappa$. The energy eigenvalues and corresponding eigenfunctions are obtained in the framework of asymptotic iteration method (AIM). Some numerical results are obtained in the presence and absence of EDC and CLT potentials.

1. Introduction

In the framework of the Dirac equation, the pseudospin (p-spin) symmetry is usually used to feature deformed nuclei, superdeformation and to establish an effective shell-model [1-3] whereas the spin symmetry is relevant for mesons [4]. Furthermore, the spin symmetry occurs when the difference of scalar potential $S(r)$ and vector potential $V(r)$ is constant, i.e. $\Delta(r) = S(r) - V(r) = C_s$ whereas the p-spin symmetry occurs when the sum of scalar and vector potentials is constant, i.e. $\Sigma(r) = S(r) + V(r) = C_{ps}$ [5,6]. About 40 years ago, pspin concept was considered for the first time in the non-relativistic framework [7,8]. The p-spin symmetry refers to a quasi-degeneracy of single nucleon doublets with non-relativistic quantum number $(n,l,j=l+1/2)$ and $(n-1,l+2,j=l+3/2)$, where $n$, $l$ and $j$ are single nucleon radial, orbital and total angular momentum quantum numbers, respectively. The total angular momentum is $j=\tilde{l}+\tilde{s}$, where $\tilde{l}=l+1$ is a pseudo-angular momentum and $\tilde{s}$ is p-spin angular momentum [9-19]. Also, tensor potentials were introduced into the Dirac equation with the replacement $\vec{p} \to \vec{p} - iM\omega\beta.\hat{r}U(r)$ and a spin-orbit coupling is added to the Dirac Hamiltonian [20-30].

Wave equations with energy-dependent potentials occur in relativistic quantum mechanics, firstly with the Pauli–Schrödinger equation [31] and recently in the Hamiltonian formulation of the relativistic many-body problem [32-34]. Also, Energy-dependent potentials have been used as a source of nonlinear Hamiltonian evolution equations [35-38] and currently applied to soliton propagation [39-41]. Recently, extensive studies on the energy-dependent potentials have appeared in some recent works [42-46].

The aim of the present work is to study the Dirac equation for the attractive scalar and repulsive vector EDC potential including the CLT potential under the pspin and spin symmetric limit. We solve the relativistic equation to obtain its bound state solutions including the energy eigenvalues and the corresponding wave functions by means of the AIM [47-51].

The structure of this paper is as follows. In section 2, we briefly present the AIM. In section 3, the Dirac equation with EDC scalar and vector potentials including the CLT potential is briefly introduced. We solve the Dirac equation under p-spin and spin symmetric limits and give some numerical results too. Finally, our concluding remarks are given in section 4.

## 2. Asymptotic Iteration Method (AIM)

The AIM has been proposed in order to solve the second-order differential equations having the form

$$\frac{d^2 y(x)}{dx^2} = \lambda_0(x)\frac{dy(x)}{dx} + S_0(x)y(x), \qquad (1)$$

where $\lambda_0(x) \neq 0$ and the variables $\lambda_0(x)$ and $S_0(x)$ are sufficiently differentiable functions [47-51]. The differential equation (1) has a general solution given as follows

$$y_n^{k+1}(x) = N\exp\left(\int^x \frac{S_n(x')}{\lambda_n(x')}dx'\right), \tag{2}$$

where $N$ is the integration constant and

$$\lambda_n(x) = \frac{d\lambda_{n-1}(x)}{dx} + S_{n-1}(x) + \lambda_0(x)\lambda_{n-1}(x), \tag{3}$$

and

$$S_n(x) = \frac{dS_{n-1}(x)}{dx} + S_0(x)\lambda_{n-1}(x), \tag{4}$$

for some $k > 0$

$$\frac{S_k(x)}{\lambda_k(x)} = \frac{S_{k-1}(x)}{\lambda_{k-1}(x)} = \alpha(x), \quad k = 1,2,3,\cdots. \tag{5}$$

The energy eigenvalues are determined by the quantization condition given by

$$\Delta_k(x) = \begin{vmatrix} \lambda_k(x) & \lambda_{k-1}(x) \\ S_k(x) & S_{k-1}(x) \end{vmatrix} = 0, \quad k = 1,2,3,\cdots, \tag{6}$$

where $k$ is the iteration number [47-51]. By using Eq. (6) and Eq. (2), one can obtain energy eigenvalues and the corresponding wave functions, respectively.

## 3. Dirac Equation Including Tensor Like Coupling Potential

The Dirac equation with scalar $S(r)$ and vector $V(r)$ potentials including a CLT interaction field to describe spin-$1/2$ particles (in units $\hbar = c = 1$) has the explicit form:

$$\left[\vec{\alpha}\cdot\vec{p} + \beta(M + S(r)) - i\beta\vec{\alpha}\cdot\hat{r}U(r)\right]\psi(\vec{r}) = \left[E_{n\kappa} - V(r)\right]\psi(\vec{r}), \tag{7}$$

where $E_{n\kappa}$ is the relativistic energy of the quantum system, $\vec{p} = -i\vec{\nabla}$ is the three-dimensional momentum operator and $M$ is the fermionic mass. Further, $\vec{\alpha}$ and $\beta$ are the $4\times 4$ usual Dirac matrices.

Following the procedures of derivation described by Eq. (6) to Eq. (11) of Ref. [30], we finally arrive at two Schrödinger-like second-order differential equations for the upper and lower radial spinor components

$$\left[\frac{d^2}{dr^2} - \frac{\kappa(\kappa+1)}{r^2} + \frac{2\kappa}{r}U(r) - \frac{dU(r)}{dr} - U^2(r) - \frac{\left(\frac{dM}{dr} - \frac{d\Delta(r)}{dr}\right)}{M + E_{n\kappa} - \Delta(r)}\left(\frac{d}{dr} + \frac{\kappa}{r} - U(r)\right)\right]F_{n\kappa}(r)$$

$$= \left[ \left( M + E_{n\kappa} - \Delta(r) \right) \left( M - E_{n\kappa} + \Sigma(r) \right) \right] F_{n\kappa}(r), \qquad (8)$$

and

$$\left[ \frac{d^2}{dr^2} - \frac{\kappa(\kappa-1)}{r^2} + \frac{2\kappa}{r} U(r) + \frac{dU(r)}{dr} - U^2(r) - \frac{\left( \frac{dM}{dr} + \frac{d\Sigma(r)}{dr} \right)}{M - E_{n\kappa} + \Sigma(r)} \left( \frac{d}{dr} - \frac{\kappa}{r} + U(r) \right) \right] G_{n\kappa}(r)$$

$$= \left[ \left( M + E_{n\kappa} - \Delta(r) \right) \left( M - E_{n\kappa} + \Sigma(r) \right) \right] G_{n\kappa}(r), \qquad (9)$$

respectively, where $\kappa(\kappa+1) = l(l+1)$ and $\kappa(\kappa-1) = \tilde{l}(\tilde{l}+1)$. The spin-orbit quantum number $\kappa$ is related to the quantum numbers $l$ ($\tilde{l}$) for spin (p-spin) symmetry as

$$\kappa = \begin{cases} -(l+1) = -\left(j+\frac{1}{2}\right) & (s_{1/2}, p_{3/2}, \cdots \text{etc}) \quad j = l + \frac{1}{2}, \text{ aligned spin}\,(\kappa < 0) \\ +l = +\left(j+\frac{1}{2}\right) & (p_{1/2}, d_{3/2}, \cdots \text{etc}) \quad j = l - \frac{1}{2}, \text{ unaligned spin}\,(\kappa > 0), \end{cases}$$

and the quasidegenerate doublet structure can be expressed in terms of a p-spin angular momentum $\tilde{s} = 1/2$ and pseudo-orbital angular momentum $\tilde{l}$, which is defined as

$$\kappa = \begin{cases} -\tilde{l} = -(j+\frac{1}{2}) & (s_{1/2}, p_{3/2}, \cdots \text{etc}) \quad j = \tilde{l} - \frac{1}{2}, \text{ alinged pspin}\,(\kappa < 0) \\ +(\tilde{l}+1) = +(j+\frac{1}{2}) & (d_{3/2}, f_{5/2}, \cdots \text{etc}) \quad j = \tilde{l} + \frac{1}{2}, \text{ unaligned pspin}\,(\kappa > 0), \end{cases}$$

where $\kappa = \pm 1, \pm 2, \ldots$. For example, $(1s_{1/2}, 0d_{3/2})$ and $(1p_{3/2}, 0f_{5/2})$ can be considered as pspin doublets.

## 3.1. P-spin Symmetry Limit

The p-spin symmetry case occurs in the Dirac equation when $\frac{d\Sigma(r)}{dr} = 0$ or $\Sigma(r) = C_{ps} = $ constant [6,52]. The difference potential $\Delta(r)$ is simply taken as EDC potential, that is,

$$\Delta(r) = V_{\text{EDC}}(r) = -\frac{C(1+\delta E_{n\kappa})}{r}, \qquad (10)$$

where $C = Z\alpha$, $\alpha = e^2$ is the fine structure constant (in units $\hbar = c = 1$) and $\delta > 0$ [43, 44]. Further, the CLT potential is added as

$$U(r) = -\frac{A}{r}, \qquad A = \frac{Z_a Z_b e^2}{4\pi\varepsilon_0}, \qquad r \geq R_c \tag{11}$$

where $R_c = 7.78\, fm$ is the Coulomb radius, $Z_a$ and $Z_b$ denote the charges of the projectile $a$ and the target nuclei $b$, respectively [28]. Therefore, from Eq. (9), one obtains

$$\left[\frac{d^2}{dr^2} - \frac{\Lambda_\kappa(\Lambda_\kappa - 1)}{r^2} + \frac{C(1+\delta E_{n\kappa})\tilde{\gamma}}{r} - \tilde{\beta}^2\right] G_{n\kappa}(r) = 0, \tag{12}$$

where

$$\Lambda_\kappa = \kappa + A,$$

$$\tilde{\gamma} = E_{n\kappa} - M - C_{ps},$$

$$\tilde{\beta}^2 = (E_{n\kappa} + M)(M - E_{n\kappa} + C_{ps}). \tag{13}$$

By using the particular transformation $G_{n\kappa}(r) = r^{\tilde{\varepsilon}+1/2} e^{-\tilde{\beta}r} g_{nk}(r)$ [48], we can recast (12) as

$$\frac{d^2 g_{n\kappa}(r)}{dr^2} = \left(2\tilde{\beta} - \frac{2\tilde{\varepsilon}+1}{r}\right)\frac{dg_{n\kappa}(r)}{dr} + \left(\frac{2\tilde{\varepsilon}\tilde{\beta} + \tilde{\beta} - \tilde{\Gamma}^2}{r}\right) g_{n\kappa}(r), \tag{14}$$

where

$$\tilde{\varepsilon} = \kappa + A - \frac{1}{2}, \tag{15a}$$

$$\tilde{\Gamma}^2 = C(1+\delta E_{n\kappa})\tilde{\gamma}. \tag{15b}$$

The comparison of Eq. (14) with Eq. (1) gives the two variables

$$\lambda_0(r) = 2\tilde{\beta} - \frac{2\tilde{\varepsilon}+1}{r}, \tag{16a}$$

$$S_0(r) = \frac{2\tilde{\varepsilon}\tilde{\beta} + \tilde{\beta} - \tilde{\Gamma}^2}{r}. \tag{16b}$$

Further, the substitution of Eq. (16) into Eq. (3) and Eq. (4) gives the parameters

$$\lambda_1(r) = 4\tilde{\beta}^2 - \frac{6\tilde{\beta}\tilde{\varepsilon} + 3\tilde{\beta} + \tilde{\Gamma}^2}{r} + \frac{4\tilde{\varepsilon}^2 + 6\tilde{\varepsilon} + 2}{r^2},$$

$$S_1(r) = \frac{4\tilde{\varepsilon}\tilde{\beta}^2 + 2\tilde{\beta}^2 - 2\tilde{\beta}\tilde{\Gamma}^2}{r} + \frac{2\tilde{\Gamma}^2 + 2\tilde{\varepsilon}\tilde{\Gamma}^2 - 6\tilde{\varepsilon}\tilde{\beta} - 2\tilde{\beta} - 4\tilde{\varepsilon}^2\tilde{\beta}}{r^2}. \tag{17}$$

To find energy eigenvalue, we substitute Eq. (17) into Eq. (6) and obtain

$$k=1 \Rightarrow \Delta_1(r)=0 \Rightarrow \lambda_1(r)S_0(r)-\lambda_0(r)S_1(r)=0 \Rightarrow \tilde{\varepsilon}_0 = \frac{\tilde{\Gamma}^2}{2\tilde{\beta}}-\frac{1}{2},$$

$$k=2 \Rightarrow \Delta_2(r)=0 \Rightarrow \lambda_2(r)S_1(r)-\lambda_1(r)S_2(r)=0 \Rightarrow \tilde{\varepsilon}_1 = \frac{\tilde{\Gamma}^2}{2\tilde{\beta}}-\frac{3}{2},$$

.
.
.

etc. (18)

and generally for arbitrary $k$, we have

$$\tilde{\varepsilon}_n = \frac{\tilde{\Gamma}^2}{2\tilde{\beta}}-\left(n+\frac{1}{2}\right). \tag{19}$$

Recalling Eqs. (13) and (15) and substituting in above equation, the energy eigenvalues function is obtained as

$$4(M+E_{n\kappa})(n+\kappa+A)^2 = C^2(M-E_{n\kappa}+C_{ps})(1+\delta E_{n\kappa})^2, \tag{20}$$

when $A=\delta=C_{ps}=0$, the problem reduces to the pure Coulomb potential and the energy eigenvalues of p-spin symmetry are obtained as [26]

$$E_{n\kappa}^{\text{Coulomb}} = -M\frac{(n+\kappa)^2-Z^2\alpha^2}{(n+\kappa)^2+Z^2\alpha^2}. \tag{21}$$

Obviously, the above energy formula is the same as Eq. (40) of Ref. [53] in Klein-Gordon-Coulomb problem $(\kappa=l+1, C=q)$ corresponding to antiparticle case under the condition of $S(r)=-V(r)$ or $\Sigma(r)=0$ (exact p-spin symmetry, $C_{ps}=0$) whereas the particle has continuous solution for all states $E_{n\kappa}=M$.

Some numerical results of Eqs. (20) and (21) are given in tables 1. We use parameters as: $M=5.0\,fm^{-1}$, $C=0.5$, $C_{ps}=0$, $A=1$ [26] and $\delta=0.05$. In the limiting case when $A=\delta=0$, we see that bound state energies $(1s_{1/2},0d_{3/2})$, $(1p_{3/2},0f_{5/2})$, $(1d_{5/2},0g_{7/2})$ and $(1f_{7/2},0h_{9/2})$, …etc. are becoming degenerate, where each pair is considered as p-spin doublet. Further, in the presence or tensor potential, $A\neq 0$ and $\delta=0$, the degeneracies are removed and the energy levels of the p-spin aligned states and p-spin unaligned states move in the opposite directions. For example: in p-spin doublet $(1s_{1/2},0d_{3/2})$; when $A=\delta=0$, $E_{1,-1}=E_{1,2}=-4.931034483\,fm^{-1}$, but when $A=1$ and $\delta=0$,

$E_{1,-1} = -4.846153846 fm^{-1}$ with $\kappa < 0$ and $E_{1,2} = -4.961089494 fm^{-1}$ with $\kappa > 0$. As wee can see from table 1, the energy-dependent potential can not alone remove such degeneracy.

In our calculations to wave functions, we use Eq. (2) and obtain

$$g_0(r) = N,$$

$$g_1(r) = N(\tilde{\Gamma}^2 - 2\tilde{\beta})\left(1 - \frac{2\tilde{\beta}^2}{\tilde{\Gamma}^2 - 2\tilde{\beta}}\right),$$

.

.

etc. (22)

which finally leads to the form

$$G_{n\kappa}(r) = B_n r^{\tilde{\varepsilon}_n + 1/2} e^{-\tilde{\beta} r} (-1)^n \left(\prod_{k=n}^{2n-1} (\tilde{\Gamma}^2 - (k+1)\tilde{\beta})\right) {}_1F_1(-n, 2\tilde{\varepsilon}_n + 1; 2\tilde{\beta} r), \quad (23)$$

where $B_n$ is new normalization constant. We can also express the hypergeometric function ${}_1F_1$ in terms of Laguerre polynomials as

$$G_{n\kappa}(r) = D_{n\kappa} r^{\tilde{\varepsilon}_n + 1/2} e^{-\tilde{\beta} r} L_n^{2\tilde{\varepsilon}_n}(2\tilde{\beta} r), \quad (24)$$

where $\tilde{D}_{n\kappa}$ is normalization constant given as [54]

$$\tilde{D}_{n\kappa} = \frac{1}{n!} (2\tilde{\beta})^{\tilde{\varepsilon}_n + 1/2} \sqrt{\frac{(n - 2\tilde{\varepsilon}_n)!}{n!}}. \quad (25)$$

The upper spinor component of the Dirac equation can be calculated as [30]

$$F_{n\kappa}(r) = \frac{1}{M - E_{n\kappa} + C_{ps}} \left(\frac{d}{dr} - \frac{\kappa}{r} + U(r)\right) G_{n\kappa}(r), \quad (26)$$

where $E_{n\kappa} \neq M$ when $C_{ps} = 0$ (exact pspin symmetry) which means that only negative energy solutions are permissible.

## 3.2. Spin Symmetry Limit

In the spin symmetric limit $\frac{d\Delta(r)}{dr} = 0$ or $\Delta(r) = C_s = $ constant [6,55], then Eq. (11) with $\Sigma(r)$ as EDC potential and including Coulomb-like tensor potential becomes

$$\left[\frac{d^2}{dr^2} - \frac{\eta_\kappa(\eta_\kappa - 1)}{r^2} + \frac{\gamma C(1+E_{n\kappa})}{r} - \beta^2\right] F_{n\kappa}(r) = 0, \qquad (27)$$

where $\gamma = M + E_{n\kappa} - C_s$ and $\beta^2 = (M - E_{n\kappa})(M + E_{n\kappa} - C_s)$, $\eta_\kappa = \kappa + A + 1$ and also, $\kappa = l$ and $\kappa = -l-1$ for $\kappa < 0$ and $\kappa > 0$, respectively. To avoid repetition, by following the procedure of previous subsection, the energy eigenvalue equation of spin symmetry is obtained as

$$4(M - E_{n\kappa})(n + \kappa + A)^2 = C^2(M + E_{n\kappa} - C_s)(1 + \delta E_{n\kappa})^2, \qquad (28)$$

when $A = \delta = C_s = 0$, the problem reduces into the pure Coulomb potential and the energy eigenvalues of spin symmetry are obtained as

$$E_{n\kappa}^{Coulomb} = M \frac{(n+\kappa+1)^2 - Z^2\alpha^2}{(n+\kappa+1)^2 + Z^2\alpha^2}. \qquad (29)$$

Obviously, the above energy formula is the same as Eq. (38) of Ref. [53] in Klein-Gordon-Coulomb problem $(\kappa = l, C = q)$ corresponding to particle energy solution under the condition of $S(r) = V(r)$ or $\Delta(r) = 0$ (exact spin symmetry, $C_s = 0$) whereas the antiparticle has continuous solution for all states $E_{n\kappa} = -M$.

Some numerical results of Eqs. (28) and (29) are given in table 2. We use same parameters as previous subsection. We can observe that every pair of orbitals ($np_{1/2}, np_{3/2}$), ($nd_{3/2}, nd_{5/2}$) and ($nf_{5/2}, nf_{7/2}$) has the same energy in the absence of the tensor potential ($A = 0$). Thus, they can be viewed as the spin doublets, i.e., the state $1p_{1/2}$ with $n = 1$ and $\kappa = 1$ forms a spin doublet with the $1p_{3/2}$ state with $n = 1$ and $\kappa = -2$. On the other hand, in the presence of the tensor potential ($A \neq 0$), one can notice that degeneracy between every pair of spin doublets is removed.

In the presence or tensor potential, $A \neq 0$ and $\delta = 0$, the degeneracies are removed and the energy levels of the spin aligned states and spin unaligned states move in the same directions. For example: in p-spin doublet $(0p_{3/2}, 0p_{1/2})$; when $A = \delta = 0$, $E_{0,-2} = E_{0,1} = 4.846153846 fm^{-1}$, but when $A = 1$ and $\delta = 0$, $E_{0,-2} = 4.411764706 fm^{-1}$ with $\kappa < 0$ and $E_{1,2} = 4.931034483 fm^{-1}$ with $\kappa > 0$. Again, as wee can see from table 2, the energy-dependent potential can not alone remove such degeneracy.

Also, for the wave function, we obtain

$$F_{n\kappa}(r) = D_{n\kappa} r^{\varepsilon_{n\kappa}+\frac{1}{2}} e^{-\beta r} L_n^{2\varepsilon_{n\kappa}}(2\beta r), \qquad (30)$$

where $\varepsilon_{n\kappa} = \kappa + A + \dfrac{1}{2}$ and $D_{n\kappa} = \dfrac{1}{n!}(2\beta)^{\varepsilon_{n\kappa}+1/2}\sqrt{\dfrac{(n-2\varepsilon_{n\kappa})!}{n!}}$ is normalization constant.

The lower component of the Dirac spinor can be calculated as [30]

$$G_{n\kappa}(r) = \dfrac{1}{M + E_{n\kappa} - C_s}\left(\dfrac{d}{dr} + \dfrac{\kappa}{r} - U(r)\right)F_{n\kappa}(r), \qquad (31)$$

where $E \neq -M + C_s$.

## 4. Concluding Remarks

We investigated the Dirac problem with EDC potential including coupling CLT interaction potential under p-spin and spin symmetry limits. We have calculated analytical expressions for energy eigenvalue equations and normalized wave functions by using the AIM. We have also shown that the tensor interaction removes degeneracies between pspin doublets. In tables 1 and 2, we have presented some numerical results in the presence and absence of EDC potential and CLT potential. The bound-state energy solution of the Dirac-Coulomb problem with p-spin (spin) symmetry resembles the Klein-Gordon bound-state energy of the antiparticle (particle) when $S(r) = -V(r)$ ($S(r) = V(r)$) case.

## Acknowledgments

We thank the referee(s) for their invaluable suggestions and comments that have greatly helped in improving this paper.

## References

[1] A. Bohr, I. Hamamoto and B. R. Mottelson, Phys. Scr. **26** (1982) 267.

[2] J. Dudek, W. Nazarewicz, Z. Szymanski and G. A. Leander, Phys. Rev. Lett. **59** (1987) 1405.

[3] D. Troltenier, C. Bahri and J. P. Draayer, Nucl. Phys. A **586** (1995) 53.

[4] P. R. Page, T. Goldman and J. N. Ginocchio, Phys. Rev. Lett. **86** (2001) 204.

[5] J. N. Ginocchio, A. Leviatan, J. Meng, and S. G. Zhou, Phys. Rev. C **69** (2004) 034303.

[6] J. N. Ginocchio, Phys. Rev. Lett. **78** (3) (1997) 436.

[7] K. T. Hecht and A. Adler, Nucl. Phys. A **137** (1969) 129.

[8] A. Arima, M. Harvey and K. Shimizu, Phys. Lett. B **30** (1969) 517.

[9] B. D. Serot, Rep. Prog. Phys. **55** (1992) 1855.

[10] D. Troltenier, C. Bahri and J. P. Draayer, Nucl. Phys. A **586** (1995) 53.

[11] J. N. Ginocchio and A. Leviatan, Phys. Lett. B **425** (1998) 1.

[12] J. Meng, K. Sugawara-Tanabe, S. Yamaji, P. Ring and A. Arima, Phys. Rev. C **58** (2) (1998) R628.

[13] P. R. Page, T. Goldman and J. N. Ginocchio, Phys. Rev. Lett. **86** (2001) 204.

[14] J. N. Ginocchio, Phys. Rep. **414** (2005) 165.

[15] D. Troltenier, C. Bahri and J. P. Draayer, Nucl. Phys. A **586** (1995) 53.

[16] A. de Souza Dutra and M. Hott, Phys. Lett. A **356** (2006) 215.

[17] A. D. Alhaidari, H. Bahlouli and A. Al-Hasan, Phys. Lett. A **349** (2006) 87.

[18] G. F. Wei and S. H. Dong, EPL **87** (2009) 40004.

[19] G. F. Wei and S. H. Dong, Phys. Lett. B **686** (2010) 288.

[20] M. Moshinsky and A. Szczepaniak, J. Phys. A: Math. Gen. **22** (1989) L817.

[21] R. F. Furnstahl, J. J. Rusnak and B. D. Serot, Nucl. Phys. A **632** (1998) 607.

[22] V. I. Kukulin, G. Loyola and M. Moshinsky, Phys. Lett. A **158** (1991) 19.

[23] G. Mao, Phys. Rev. C **67** (2003) 044318.

[24] R. Lisboa, M. Malheiro, A.S. de Castro, P. Alberto and M. Fiolhais, Phys. Rev. C **69** (2004) 0243319.

[25] P. Alberto, R. Lisboa, M. Malheiro and A. S. de Castro, Phys. Rev. C **71**(2005) 034313.

[26] M. Hamzavi, A. A. Rajabi and H. Hassanabadi, Phys. Lett A **374** (2010) 4303.

[27] H. Akcay, Phys. Lett. A **373** (2009) 616.

[28] S. M. Ikhdair and R. Sever, Appl. Math. Comput. **216** (2010) 911.

[29] O. Aydoğdu and R. Sever, Few-Body Syst. **47** (2010) 193.

[30] M. Hamzavi, A. A. Rajabi and H. Hassanabadi, Few-Body Syst. **48** (2010) 171.

[31] W. Pauli, Z. Phys. **43** (1927) 601.

[32] V. A. Rizov, H. Sazdjian and I. T. Todorov , Ann. Phys.(NY) **165** (1985) 59.

[33] H. Sazdjian, Phys. Rev. D **33** (1986) 3401.

[34] H. Sazdjian and J. Mourad,  J. Math. Phys. **35** (1994) 6379.

[35] Alonso L. Martinez, J. Math. Phys. **21** (1980) 2342.

[36] V. E. Zakharov and A. B. Shabat, Zh. Eksp. Teor. Fiz. **61** (1971) 118.

[37] M. Jaulent and C. Jean, Commun. Math. Phys. **28** (1972) 177.

[38] M. Wadati and T. Kamijo, Prog. Theor. Phys. **52** (1974) 397.

[39] V. N. Serkin and A. Hasegawa, Phys. Rev. Lett. **85** (2000) 4502.

[40] L. F. Mollenauer, R. H. Stolen and J. P. Gordon, Phys. Rev. Lett. **45** (1980) 1095.

[41] J. R. Taylor, ed. Optical Solitons—Theory and Experiments Studies in Modern Optics (Cambridge: Cambridge University Press, 1992).

[42] J. Formanek, J. Mareš and R. J. Lombard, Czech J. Phys. **54** (2004) 289.

[43] R. J. Lombard, J. Mareš and C. Volpe, J. Phys. G: Nucl. Part. Phys. **34** (2007) 1.

[44] R. Yekken and R. J. Lombard, J. Phys. A: Math. Theor. **43** (2010) 125301.

[45] H. Hassanabadi, S. Zarrinkamar, and A.A. Rajabi, Commun. Theor. Phys. **55** (2011) 541.

[46] H. Hassanabadi, S. Zarrinkamar, H. Hamzavi and A. A. Rajabi, Arab. J. Sci. Eng. **37** (2012) 209.

[47] B. Champion, R. L. Hall, N. Saad, Int. J. Mod. Phys. A **23** (2008) 1405.

[48] H. Ciftci, R. L. Hall and N. Saad, J. Phys. A: Math. Gen. **36** (2003) 11807.

[49] F. M. Fernandez, J. Phys. A: Math. Gen. **37** (2004) 6173.

[50] H. Ciftci, R. L. Hall and N. Saad, J. Phys. A: Math. Gen. **38** (2005) 1147.

[51] H. Ciftci, R. L. Hall and N. Saad, Phys. Lett. A **340** (2005) 388.

[52] G. F. Wei and S. H. Dong, Eur. Phys. J. A **46** (2010) 207.

[53] S. M. Ikhdair, Eur. J. Phys. A **40** (2009) 143.

[54] O. Aydoğdu and R. Sever, Ann. Phys. **325** (2010) 373.

[55] G. F. Wei and S. H. Dong, Phys. Lett. A **373** (2008) 49; G. F. Wei and S. H. Dong, Phys. Lett. A **373** (2009) 2428; G. F. Wei and S. H. Dong, Eur. Phys. J. A **43** (2010) 185; G. F. Wei and S. H. Dong, Phys. Scr. **81** (2010) 035009.

**Table 1:** The bound state energy eigenvalues of the Coulomb potential (in units of $fm^{-1}$) under p-spin symmetry for various values of $n$ and $\kappa$ in the presence or absence of energy-dependence and tensor potentials.

| $\tilde{l}$ | $n,\kappa<0$ $(l,j)$ | $E_{n,\kappa<0}$ $A=0$ $\delta=0$ | $E_{n,\kappa<0}$ $A=1$ $\delta=0.05$ | $E_{n,\kappa<0}$ $A=0$ $\delta=0.05$ | $E_{n,\kappa<0}$ $A=1$ $\delta=0$ | $n-1,\kappa>0$ $(l+2,j+1)$ | $E_{n-1,\kappa>0}$ $A=0$ $\delta=0$ | $E_{n-1,\kappa>0}$ $A=1$ $\delta=0.05$ | $E_{n-1,\kappa>0}$ $A=0$ $\delta=0.05$ | $E_{n-1,\kappa>0}$ $A=1$ $\delta=0$ |
|---|---|---|---|---|---|---|---|---|---|---|
| 1 | 1, -1 $1s_{1/2}$ | −4.931034483 | −4.911857241 | −4.960887104 | −4.846153846 | 0, 2 $0d_{3/2}$ | −4.931034483 | −4.978011333 | −4.960887104 | −4.961089494 |
| 2 | 1, -2 $1p_{3/2}$ | −4.961089494 | −4.960887104 | −4.978011333 | −4.931034483 | 0, 3 $0f_{5/2}$ | −4.961089494 | −4.985930930 | −4.978011333 | −4.975062344 |
| 3 | 1, -3 $1d_{5/2}$ | −4.975062344 | −4.978011333 | −4.985930 93 | −4.961089494 | 0, 4 $0g_{7/2}$ | −4.975062344 | −4.990231203 | −4.985930 93 | −4.982668977 |
| 4 | 1, -4 $1f_{7/2}$ | −4.982668977 | −4.985930930 | −4.990231 20 | −4.975062344 | 0, 5 $0h_{9/2}$ | −4.982668977 | −4.992823542 | −4.990231 20 | −4.987261146 |
| 1 | 2, -1 $2s_{1/2}$ | −4.961089494 | −4.960887104 | −4.978011333 | −4.931034483 | 1, 2 $1d_{3/2}$ | −4.961089494 | −4.985930930 | −4.978011333 | −4.975062344 |
| 2 | 2, -2 $2p_{3/2}$ | −4.975062344 | −4.978011333 | −4.985930930 | −4.961089494 | 1, 3 $1f_{5/2}$ | −4.975062344 | −4.990231203 | −4.985930930 | −4.982668977 |
| 3 | 2, -3 $2d_{5/2}$ | −4.982668977 | −4.985930930 | −4.990231203 | −4.975062344 | 1, 4 $1g_{7/2}$ | −4.982668977 | −4.992823542 | −4.990231203 | −4.987261146 |
| 4 | 2, -4 $2f_{7/2}$ | −4.987261146 | −4.990231203 | −4.992823542 | −4.982668977 | 1, 5 $1h_{9/2}$ | −4.987261146 | −4.994505831 | −4.992823542 | −4.990243902 |

**Table 2:** The bound state energy eigenvalues of the Coulomb potential (in unit of $fm^{-1}$) with spin symmetry for several values of $n$ and $\kappa$ in the presence or absence of energy-dependence and tensor potentials.

| $l$ | $n, \kappa < 0$ ($l, j = l+1/2$) | $E_{n,\kappa<0}$ $A=0$ $\delta=0$ | $E_{n,\kappa<0}$ $A=1$ $\delta=0.05$ | $E_{n,\kappa<0}$ $A=0$ $\delta=0.05$ | $E_{n,\kappa<0}$ $A=1$ $\delta=0$ | $n-1, \kappa > 0$ ($l, j = l-1/2$) | $E_{n-1,\kappa>0}$ $A=0$ $\delta=0$ | $E_{n-1,\kappa>0}$ $A=1$ $\delta=0.05$ | $E_{n-1,\kappa>0}$ $A=0$ $\delta=0.05$ | $E_{n-1,\kappa>0}$ $A=1$ $\delta=0$ |
|---|---|---|---|---|---|---|---|---|---|---|
| 1 | 0, -2 $0p_{3/2}$ | 4.846153846 | 4.163939478 | 4.766014138 | 4.411764706 | 0, 1 $0p_{1/2}$ | 4.846153846 | 4.893560178 | 4.766014138 | 4.931034483 |
| 2 | 0, -3 $0d_{5/2}$ | 4.931034483 | 4.766014138 | 4.893560178 | 4.846153846 | 0, 2 $0d_{3/2}$ | 4.931034483 | 4.939625999 | 4.893560178 | 4.961089494 |
| 3 | 0, -4 $0f_{7/2}$ | 4.961089494 | 4.893560178 | 4.939625999 | 4.931034483 | 0, 3 $0f_{5/2}$ | 4.961089494 | 4.961209681 | 4.939625999 | 4.975062344 |
| 4 | 0, -5 $0g_{9/2}$ | 4.975062344 | 4.939625999 | 4.961209681 | 4.961089494 | 0, 4 $0g_{7/2}$ | 4.975062344 | 4.973004886 | 4.961209681 | 4.982668977 |
| 1 | 1, -2 $1p_{3/2}$ | 4.931034483 | 4.766014138 | 4.893560178 | 4.846153846 | 1, 1 $1p_{1/2}$ | 4.931034483 | 4.939625999 | 4.893560178 | 4.961089494 |
| 2 | 1, -3 $1d_{5/2}$ | 4.961089494 | 4.893560178 | 4.939625999 | 4.931034483 | 1, 2 $1d_{3/2}$ | 4.961089494 | 4.961209681 | 4.939625999 | 4.975062344 |
| 3 | 1, -4 $1f_{7/2}$ | 4.975062344 | 4.939625999 | 4.961209681 | 4.961089494 | 1, 3 $1f_{5/2}$ | 4.975062344 | 4.973004886 | 4.961209681 | 4.982668977 |
| 4 | 1, -5 $1g_{9/2}$ | 4.982668977 | 4.961209681 | 4.973004886 | 4.975062344 | 1, 4 $1g_{7/2}$ | 4.982668977 | 4.980141318 | 4.973004886 | 4.987261146 |